\documentclass[twocolumn,floatfix,eqsecnum,rmp,superscriptaddress]{revtex4}
\usepackage{graphicx}
\usepackage{amsmath}
\usepackage{bm}
\bibpunct{[}{]}{,}{n}{}{,}
\newcommand{\be}{\begin{equation}}
\newcommand{\ee}{\end{equation}}

\begin{document}
\title{The Superconducting Quantum Point Contact}
\author{C. W. J. Beenakker}
\affiliation{Philips Research Laboratories, 5600 JA Eindhoven, The Netherlands}
\affiliation{Institute for Theoretical Physics, University of California, Santa Barbara, CA 93106,
USA}
\author{H. van Houten}
\affiliation{Philips Research Laboratories, 5600 JA Eindhoven, The Netherlands}
\begin{abstract}
The theoretical prediction for the
discretization of the supercurrent in a quantum point contact is
reviewed in detail. A general relation is derived between the
supercurrent and the quasiparticle excitation spectrum of a Josephson
junction.  The discrete spectrum of a quantum point contact between two
superconducting reservoirs with phase difference
$\delta\phi\in(-\pi/2,\pi/2)$ is shown to consist of a multiply
degenerate state at energy $\Delta_{0}\cos(\delta\phi/2)$ (one state
for each of the $N$ propagating modes at the Fermi energy).  The
resulting zero-temperature supercurrent is
$I=N(e\Delta_{0}/\hbar)\sin(\delta\phi/2)$. The critical current is
discretized in units of $e\Delta_{0}/\hbar$, dependent on the energy
gap $\Delta_{0}$ of the bulk superconductor but not on the junction
parameters. To achieve this analogue of the conductance quantization in
the normal state, it is essential that the constriction is short
compared to the superconducting coherence length.\bigskip\\
{\tt Published in {\em Nanostructures and Mesoscopic Systems},\\ 
edited by W. P. Kirk and M. A. Reed (Academic, New York, 1992).}
\end{abstract}
\maketitle

\tableofcontents

\section{\label{sec1} Introduction}

The work reviewed in this contribution was motivated by a sequence of
analogies. First came the quantum point contact (QPC) for {\em
electrons} \cite{Wee88,Wha88}, a constriction in a two-dimensional
electron gas with a quantized conductance. Then followed the optical
analogue: The predicted \cite{Hou90} discretization of the transmission
cross-section of a diffusely illuminated aperture or slit was recently
observed experimentally \cite{Mon91}. One can speak of this optical
analogue as a QPC for {\em photons}. Can one extend the analogue
towards a QPC for {\em Cooper pairs\/}? The answer is {\em Yes}
\cite{Bee91}.  A narrow and short, impurity-free constriction in a
superconductor has a zero-temperature critical current which is an
integer multiple of $e\Delta_{0}/\hbar$, with $\Delta_{0}$ the energy
gap of the bulk superconductor. In this paper we present our theory for
the discretization of the supercurrent, in more detail than we could in
our original publication \cite{Bee91}. For a less detailed, but more
tutorial presentation, see Ref.\ \cite{Len91}. To introduce the reader
to the problem, we first summarize some basic facts on the Josephson
effect in classical point contacts \cite{Lik79}.

The theory of the stationary Josephson effect for a
classical ballistic point contact is due to Kulik and Omel'yanchuk
\cite{Kul77}. The adjectives {\em classical\/} and {\em ballistic\/}
refer to the regime $\lambda_{\rm F}\ll W\ll l$, where $\lambda_{\rm
F}$ is the Fermi wavelength, $W$ the width of the point contact, and
$l$ the mean free path. The width $W$ and length $L$ of the
constriction are also assumed to be much smaller than the
superconducting coherence length $\xi_{0}=\hbar v_{\rm
F}/\pi\Delta_{0}$ (with $v_{\rm F}$ the Fermi velocity).  It is then
irrelevant for the Josephson effect whether the constriction is made
out of a superconductor or a normal metal. Kulik and Omel'yanchuk
calculated the relationship between the supercurrent $I$ and the phase
difference $\delta\phi$ of the pair potential in the superconducting
reservoirs at opposite sides of the constriction. Their
zero-temperature result is \begin{eqnarray} I=\pi
G\frac{\Delta_{0}}{e}\sin(\delta\phi/2), \;|\delta\phi|<\pi ,\label{KO}
\end{eqnarray} where $G$ is the normal-state conductance of the point
contact. For a three-dimensional (3D) point contact of cross-sectional
area $S$ one has $G=(2e^{2}/h)k_{\rm F}^{2}S/4\pi$, with $k_{\rm F}$
the Fermi wavevector. (A 2D point contact of width $W$ has
$G=(2e^{2}/h)k_{\rm F}W/\pi$.) These conductances, which are
independent of the mean free path for $W\ll l$, are {\em contact\/}
conductances. The current-phase relationship (\ref{KO}) is plotted in
Fig.\ 1 (solid curve).

\begin{figure}
\centerline{\includegraphics[width=8cm]{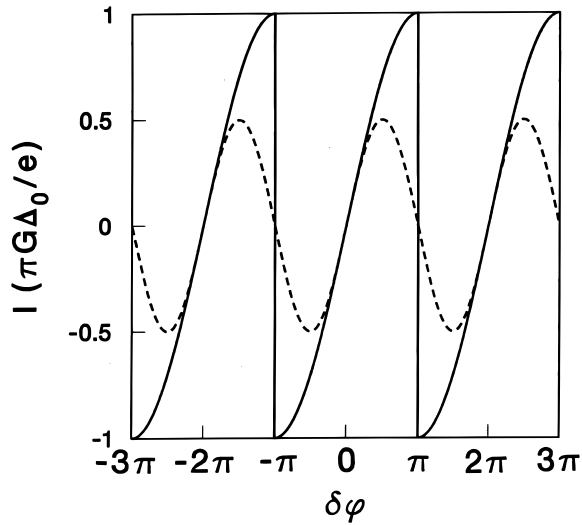}}
\caption{
Current-phase relationship at $T=0$ of a classical ballistic
point contact, according to Eq.\ (\protect\ref{KO}) (solid curve), and
of a tunnel junction, according to Eq.\ (\protect\ref{AB}) (dashed
curve).
\label{fig1}
}
\end{figure}

As always, the Josephson current is an odd
function of $\delta\phi$, and periodic with period $2\pi$. The critical
current $I_{\rm c}\equiv{\rm max}\,I(\delta\phi)$ is reached at
$\delta\phi=\pi$, and equals $I_{\rm c}=\pi G\Delta_{0}/e$. At the
critical current, the $I(\delta\phi)$ curve is discontinuous for $T=0$.
This discontinuity is smoothed out at finite temperatures.

The supercurrent in the case of a ballistic point contact has a
markedly different dependence on the phase than in the case of a tunnel
junction, for which the Josephson effect was first proposed
\cite{Jos62}. The $T=0$ current-phase relationship for a tunnel
junction is given by \cite{And63,Amb63} \begin{eqnarray}
I={\textstyle\frac{1}{2}}\pi
G\frac{\Delta_{0}}{e}\sin(\delta\phi),\label{AB} \end{eqnarray} plotted
as a dashed curve in Fig.\ 1. There is now no discontinuity in
$I(\delta\phi)$. The critical current $I_{\rm c}=\pi G\Delta_{0}/2e$ is
reached at $\delta\phi=\pi/2$. In both cases, $I_{\rm c}$ is
proportional to $G$, but with different proportionality coefficients.
The ratio $I_{\rm c}/G$ is therefore not universal.

Point contacts and tunnel junctions have in common that the length of
the junction or weak link is much smaller than $\xi_{0}$. At low
temperatures, the Josephson effect can occur also in junctions much
longer than $\xi_{0}$ \cite{Asl68,Kul69}. Consider an SNS junction
(S=super\-con\-duc\-tor, N=normal metal) having a length $L_{\rm N}$ of
the normal-metal region (i.e.\ the separation of the two SN interfaces)
which greatly exceeds the coherence length $\xi_{0}$ of the
superconductors (while still $L_{\rm N}\ll l$). The current-phase
relationship for the SNS junction at $T=0$ is a sawtooth
\cite{Ish70,Svi72,Bar72} \begin{eqnarray} I=\alpha G\frac{\hbar v_{\rm
F}}{eL_{\rm N}}\, \frac{\delta\phi}{\pi},\;|\delta\phi|<\pi ,
\label{SNS} \end{eqnarray} with critical current $I_{\rm c}=\alpha
G\hbar v_{\rm F}/eL_{\rm N}$ (reached at $\delta\phi=\pi$). The
numerical coefficient $\alpha$ is of order unity and depends on the
dimensionality of the junction. (There is a disagreement among
Refs.\ \cite{Ish70} and \cite{Bar72} on the precise value of $\alpha$.)
The $I_{\rm c}/G$ ratio is now a function of $L_{\rm N}$. This is in
contrast to the point contact and tunnel junction, where $I_{\rm c}/G$
depends only on the gap $\Delta_{0}$ in the bulk reservoir---not on the
geometry of the junction. There is also a different temperature
dependence. The supercurrents (\ref{KO}) and (\ref{AB}) decay when $T$
approaches the critical temperature $T_{\rm c}\simeq \Delta_{0}/k_{\rm
B}$, wheras Eq.\ (\ref{SNS}) requires temperatures smaller than
$T_{\rm c}$ by a factor $\xi_{0}/L_{\rm N}\ll 1$.

The problem addressed in Ref.\ \cite{Bee91} was to find the
superconducting analogue of the conductance quantization in the normal
state, which is a geometry-independent effect. The system in which we
looked for such an effect was a short ballistic point contact, much
shorter than $\xi_{0}$, and of a width comparable to $\lambda_{\rm
F}$.  One would hope that a quantized $G$ will lead to a
``quantization'' (or, more correctly, a {\em discretization}) of
$I_{\rm c}$ in units which will still depend on $\Delta_{0}$, but which
are independent of the properties of the junction. The long SNS
junction is not a likely candidate because of its $L_{\rm N}$-dependent
critical current. Still, if a quantum point contact is inserted in a
long SNS junction ($L_{\rm N}\gg\xi_{0}$), one might expect to find
{\em geometry-dependent\/} quantum size effects on the Josephson
current.  Theoretical work on that structure was done by Furusaki,
Takayanagi, and Tsukada \cite{Fur91}. Their result is that $I_{\rm c}$
increases stepwise with increasing width for certain shapes of the
constriction, but not generically. When steps do occur, the step height
depends sensitively on the parameters of the junction. As we will show,
a short junction (short compared to $\xi_{0}$) is essential for a
generic, junction-independent behavior.

In Sec.\ II we describe the method that we use to calculate the
Josephson current through a quantum point contact, which is different
from the method used in Ref.\ \cite{Kul77} for a classical point
contact. The central result of that section is a relation between the
supercurrent and the quasiparticle excitation spectrum of the Josephson
junction. In Sec.\ III the actual calculation is presented, following
Ref.\ \cite{Bee91}. We conclude in Sec.\ IV with a brief discussion of
a possible experimental realization of the superconducting quantum
point contact.

The present theory is based on two approximations: Firstly, we assume
adiabatic transport, i.e.\ absence of scattering between the transverse
modes (or 1D subbands) in the quantum point contact. The assumption of
adiabaticity brings out the essential features of the effect in the
simplest and most direct way, just as it does in the normal state
\cite{Gla88}. As in the normal state \cite{Yac90}, we expect the
discretization of the Josephson current to be robust to deviations from
adiabaticity which do not cause backscattering. In particular, although
we assume both $l\gg W$ and $l\gg\xi_{0}$ in the present analysis, we
believe that the condition $l\gg W$ on the mean free path is
sufficient, since scattering in the wide regions outside the
constriction is unlikely to cause backscattering into the
constriction.

Secondly, in order to further simplify the problem, we treat the
propagation of the modes in the WKB approximation. The WKB
approximation breaks down if the Fermi energy $E_{\rm F}$ lies within
$\Delta_{0}$ from the cutoff energy of a transverse mode (being the 1D
subband bottom). We can therefore only describe the plateau region of
the discretized Josephson current, not the transition from one multiple
of $e\Delta_0/\hbar$ to the next. Since the transition region is
smaller than the plateau region by a factor of order
$N\Delta_{0}/E_{\rm F}$ (where $N$ is the number of occupied subbands
in the constriction), it is relatively unimportant if $\Delta_{0}\ll
E_{\rm F}$ and $N\simeq 1$.

\section{\label{sec2} Josephson current from excitation spectrum}

The theory of Kulik and Omel'yanchuk \cite{Kul77} on the Josephson
effect in a classical point contact is based on a classical
Boltzmann-type transport equation for the Green's functions (the
Eilenberger equation \cite{Lik79}), which is not applicable to a
quantum point contact. The present analysis is based on the fully
quantum-mechanical Bogoliubov-de Gennes (BdG) equations for
quasiparticle wavefunctions, into which the Green's functions can be
expanded \cite{deG66}. In the present section we describe how the
Josephson current can be obtained directly from the quasi-particle
excitation spectrum---without having to calculate the Green's
functions. This method is particularly well suited for the point
contact Josephson junction, which has an excitation spectrum of a very
simple form.

The BdG equations consist of two Schr\"{o}dinger equations for electron
and hole wavefunctions ${\rm u}({\bf r})$ and ${\rm v}({\bf r})$,
coupled by the pair potential $\Delta({\bf r})$:  \begin{eqnarray}
\left(\begin{array}{cc} {\cal H}&\Delta\\ \Delta^{\ast}&-{\cal
H^{\ast}} \end{array}\right) \left(\begin{array}{c}{\rm u}\\{\rm
v}\end{array}\right)=\epsilon \left(\begin{array}{c}{\rm u}\\{\rm
v}\end{array}\right) ,\label{BdG1} \end{eqnarray} where ${\cal H}=({\bf
p}+e{\bf A})^{2}/2m+V-E_{\rm F}$ is the single-electron Hamiltonian in
the field of a vector potential ${\bf A}({\bf r})$ and electrostatic
potential $V({\bf r})$. The excitation energy $\epsilon$ is measured
relative to the Fermi energy. Since the matrix operator in
Eq.\ (\ref{BdG1}) is hermitian, the eigenfunctions $\Psi=({\rm u},{\rm
v})$ form a complete orthonormal set. One readily verifies that if
$({\rm u},{\rm v})$ is an eigenfunction with eigenvalue $\epsilon$,
then $(-{\rm v}^{\ast},{\rm u}^{\ast})$ is also an eigenfunction, with
eigenvalue $-\epsilon$. The complete set of eigenvalues thus lies
symmetrically around zero. The excitation spectrum consists of all
positive $\epsilon$.

In a uniform system with $\Delta({\bf r})\equiv\Delta_{0}{\rm e}^{{\rm
i}\phi}$, ${\bf A}({\bf r})\equiv 0$, $V({\bf r})\equiv 0$, the
solution of the BdG equations is \begin{eqnarray} \epsilon&=&\left(
(\hbar^{2}k^{2}/2m-E_{\rm F})^{2}+\Delta_{0}^{2}\right) ^{1/2}
,\label{uniform1}\\ {\rm u}({\bf r})&=&(2\epsilon)^{-1/2} {\rm e}^{{\rm
i}\phi/2}\left( \epsilon+ \hbar^{2}k^{2}/2m-E_{\rm F}\right) ^{1/2}
{\rm e}^{{\rm i}{\bf k}\cdot{\bf r}} ,\label{uniform2}\\ {\rm v}({\bf
r})&=&(2\epsilon)^{-1/2} {\rm e}^{-{\rm i}\phi/2}\left( \epsilon-
\hbar^{2}k^{2}/2m+E_{\rm F}\right) ^{1/2} {\rm e}^{{\rm i}{\bf
k}\cdot{\bf r}}.\label{uniform3} \end{eqnarray} The excitation spectrum
is continuous, with excitation gap $\Delta_{0}$. The eigenfunctions
$({\rm u},{\rm v})$ are plane waves characterized by a wavevector ${\bf
k}$. The coefficients of the plane waves are the two coherence factors
of the BCS theory. As we will discuss in the next section, the
excitation spectrum acquires a discrete part in the presence of a
Josephson junction. The discrete spectrum corresponds to bound states
in the gap ($\epsilon <\Delta_{0}$), localized within $\xi_{0}$ from
the junction. The Josephson current turns out to be essentially
determined by the discrete part of the excitation spectrum.

Close to the junction the pair potential is not uniform. To determine
$\Delta({\bf r})$ one has to solve the self-consistency equation
\cite{deG66} \begin{eqnarray} \Delta({\bf r})=g({\bf r})
\sum_{\epsilon>0}{\rm v}^{\ast}({\bf r}){\rm u}({\bf r})
[1-2f(\epsilon)] ,\label{selfconsist} \end{eqnarray} where the sum is
over all states with positive eigenvalue,\footnote{A cutoff at
$\hbar\omega_{\rm D}$, with $\omega_{\rm D}$ the Debije frequency, has
to be introduced as usual in the BCS theory.} and
$f(\epsilon)=[1+\exp(\epsilon/k_{\rm B}T)]^{-1}$ is the Fermi function.
The coefficient $g$ is minus the interaction constant of the BCS theory
of superconductivity. At an SN interface, $g$ changes abruptly (over
atomic distances) from a positive constant to zero. In contrast, the
Cooper pair density $\Delta/g$ goes to zero only over macroscopic
distances (on the order of $\xi_{0}$ for a planar SN interface).
Because of Eq.\ (\ref{selfconsist}), the determination of the
excitation energy spectrum is a non-linear problem, which is further
complicated by the fact that the vector potential has also to be
determined self-consistently from the current density, via Maxwell's
equations. (The electrostatic potential can usually be assumed to be
the same as in the normal state.)

Once the eigenvalue problem (\ref{BdG1}) is solved self-consistently,
one can calculate the equilibrium average of a single-electron operator
${\cal P}$ by means of the formula (cf.\ Ref.\ \cite{deG66})
\begin{eqnarray} \langle{\cal P}\rangle =2\sum_{\epsilon>0}\int\! d{\bf
r}\left( f(\epsilon){\rm u}^{\ast}{\cal P}{\rm u}+ [1-f(\epsilon)]{\rm
v}{\cal P}{\rm v}^{\ast}\right) .\label{Oaverage} \end{eqnarray} Notice
the reverse order, ${\rm u}^{\ast}{\cal P}{\rm u}$ versus ${\rm v}{\cal
P}{\rm v}^{\ast}$, and the different thermal weight factors,
$f(\epsilon)$ for electrons and $f(-\epsilon)=1-f(\epsilon)$ for holes.
The prefactor of 2 accounts for both spin directions. (It is assumed
that ${\cal P}$ does not couple to the spin.) We are interested in
particular in the equilibrium current density ${\bf j}({\bf r})$, which
is given by 
\begin{eqnarray} {\bf j}&=&-2\frac{e}{m}\sum_{\epsilon>0}{\rm
Re}\bigl( f(\epsilon){\rm u}^{\ast}({\bf p}+e{\bf A}) {\rm u}\nonumber\\
&&\mbox{}+
[1-f(\epsilon)]{\rm v}({\bf p}+e{\bf A}) {\rm v}^{\ast}\bigr)
,\label{j} \end{eqnarray} 
in accordance with Eq.\ (\ref{Oaverage}). The
Josephson current $I$ is the integral of ${\bf j}\cdot{\bf n}$ over the
cross-section of the junction (with ${\bf n}$ a unit vector
perpendicular to the cross-section).

There is an alternative (and often more convenient) way to arrive at
the total equilibrium current $I$ flowing through the Josephson
junction, which is to use the fundamental relation \cite{And63}
\begin{eqnarray}
I=\frac{2e}{\hbar}\,\frac{dF}{d\delta\phi}\label{dFdphi} \end{eqnarray}
between the Josephson current and the derivative of the free energy $F$
with respect to the phase difference. (The derivative is to be taken
without varying the vector potential.) To apply this relation we need
to know how to obtain $F$ from the BdG equations. This is somewhat
tricky, since (because of the pair potential) one can not express $F$
exclusively in terms of the excitation energies (as one can for
non-interacting particles). The required formula was derived by Bardeen
et al.\ \cite{Bar69} from the Green's function expression for $F$. Here
we present an alternative derivation, directly from the BdG equations.

Following De Gennes \cite{deG66}, we write \begin{eqnarray}
&&F=U-TS,\label{F1}\\ &&U=\langle{\cal H}\rangle +U_{\rm
int},\label{F2}\\ &&U_{\rm int}= -\int\! d{\bf r}\,|\Delta|^{2}/g
,\label{F3}\\ &&S=-2k_{\rm B}\sum_{\epsilon>0}[f\ln f+(1-f)\ln (1-f)]
.\label{F4} \end{eqnarray} The energy $U$ is the sum of the
single-particle energy $\langle{\cal H}\rangle$ [defined according to
Eq.\ (\ref{Oaverage})] and the interaction-potential energy $U_{\rm
int}$ (which is negative, since the interaction is attractive). The
entropy $S$ of the independent fermionic excitations can be rewritten
as \begin{eqnarray} S=-2k_{\rm B}\sum_{\epsilon>0}[\ln
f+(1-f)\epsilon/k_{\rm B}T].\label{F2bis} \end{eqnarray} Using
Eqs.\ (\ref{BdG1}) and (\ref{selfconsist}) we obtain for the
single-particle energy the expression 
\begin{eqnarray} \langle{\cal
H}\rangle &=&\sum_{\epsilon>0}[\epsilon f-\epsilon(1-f)]-2U_{\rm int}\nonumber\\
&&\mbox{}
+\sum_{\epsilon>0}\epsilon\int\! d{\bf r}\, (|{\rm u}|^{2}-|{\rm
v}|^{2}).\label{F5} \end{eqnarray} 
The term containing the u's and v's
is in fact independent of these eigenfunctions, as one sees from the
following sequence of equalities:  
\begin{eqnarray}
&&\sum_{\epsilon>0}\epsilon\int\! d{\bf r}\, (|{\rm u}|^{2}-|{\rm
v}|^{2})= \sum_{\epsilon>0}\int\! d{\bf r}\,( {\rm u}^{\ast}{\cal
H}{\rm u}+ {\rm v}{\cal H}{\rm v}^{\ast})\nonumber\\
&&\;\;\;={\textstyle\frac{1}{2}}\sum_{\epsilon}\int\! d{\bf r}\,( {\rm
u}^{\ast}{\cal H}{\rm u}+ {\rm v^{\ast}}{\cal H^{\ast}}{\rm
v})\nonumber\\
&&\;\;\;={\textstyle\frac{1}{2}}{\rm Tr}
\left(\begin{array}{cc} {\cal H}&0\\0&{\cal H^{\ast}}
\end{array}\right)={\rm Tr}\,{\cal H} .  \label{F6} \end{eqnarray} 
Here
we have made use of the completeness of the set of eigenfunctions
$({\rm u},{\rm v})$ when both positive and negative $\epsilon$ are
included. Collecting results, the expression for the free energy
becomes 
\begin{eqnarray} F&=&-2k_{\rm B}T\sum_{\epsilon>0}\ln\left[
2\cosh( \epsilon/2k_{\rm B}T)\right]\nonumber\\
&&\mbox{} +\int\! d{\bf r}\,|\Delta|^{2}/g
+{\rm Tr}\,{\cal H} .\label{F7} \end{eqnarray} 
This is the electronic
contribution to the free energy, without the energy stored in the
magnetic field. Eq.\ (\ref{F7}) agrees with Ref.\ \cite{Bar69}---except
for the term ${\rm Tr}\,{\cal H}$, which is absent in their expression
for the free energy.\footnote{Bardeen et al.\ \cite{Bar69} use the free
energy (including the magnetic field contribution) to determine the
self-consistent pair and vector potentials variationally:  The
self-consistent $\Delta$ and ${\bf A}$ minimize the total free energy
of the system. Since ${\rm Tr}\,{\cal H}$ depends on ${\bf A}$, it is
not immediately obvious to us that it is justified to disregard this
term in the variational calculation of ${\bf A}$, as was done in
Ref.\ \cite{Bar69}. A possible argument is that energies near the Fermi
energy do not contribute to ${\rm Tr}\,{\cal H}$, because eigenvalues
below $E_{\rm F}$ cancel those above $E_{\rm F}$. The cancellation is
not exact, however.}

The first term in Eq.\ (\ref{F7}) (the sum over $\epsilon$) can be
formally interpreted as the free energy of non-interacting electrons,
all of one single spin, in a ``semiconductor'' with Fermi level halfway
between the ``conduction band'' (positive $\epsilon$) and the ``valence
band'' (negative $\epsilon$). This familiar \cite{Tin75} semiconductor
model of a superconductor appeals to intuition, but does not give the
free energy correctly. The second term ($-U_{\rm int}$) in
Eq.\ (\ref{F7}) corrects for a double-counting of the interaction
energy in the semiconductor model. The third term (${\rm Tr}\,{\cal
H}$) cancels a divergence at large $\epsilon$ of the series in the
first term. Substituting $F$ into Eq.\ (\ref{dFdphi}), we obtain the
required expression for the Josephson current:  \begin{eqnarray}
I&=&-\frac{2e}{\hbar}\sum_{p}\tanh(\epsilon_{p}/2k_{\rm B}T)
\frac{d\epsilon_{p}}{d\delta\phi}\nonumber\\
&&\mbox{}-\frac{2e}{\hbar}2k_{\rm B}T
\int_{\Delta_{0}}^{\infty}\!d{\epsilon}\,\ln\left[ 2\cosh(
\epsilon/2k_{\rm B}T)\right] \frac{d\rho}{d\delta\phi}\nonumber\\
&&\mbox{}+\frac{2e}{\hbar}\,\frac{d}{d\delta\phi} \int\!d{\bf
r}\,|\Delta|^{2}/g ,\label{dFdphi2} \end{eqnarray} where we have
rewritten $\sum_{\epsilon>0}$ as a sum over the discrete positive
eigenvalues $\epsilon_{p}$ ($p=1,2,\ldots$), and an integration over
the continuous spectrum with density of states $\rho(\epsilon)$. The
term ${\rm Tr}\,{\cal H}$ in Eq.\ (\ref{F7}) does not depend on
$\delta\phi$, and therefore does not contribute to $I$. The term
$-U_{\rm int}$ does contribute in general. For the case of a point
contact Josephson junction considered in this paper, however, we will
show that this contribution (the third term in Eq.\ (\ref{dFdphi2}) can
be neglected relative to the contribution from the semiconductor model.
A calculation of the Josephson current from Eq.\ (\ref{dFdphi2}) then
requires only knowledge of the eigenvalues. This is in contrast to a
calculation based on Eq.\ (\ref{j}), which requires also the
eigenfunctions.

\section{\label{sec3} Josephson current through a quantum point contact}

\begin{figure}
\centerline{\includegraphics[width=8cm]{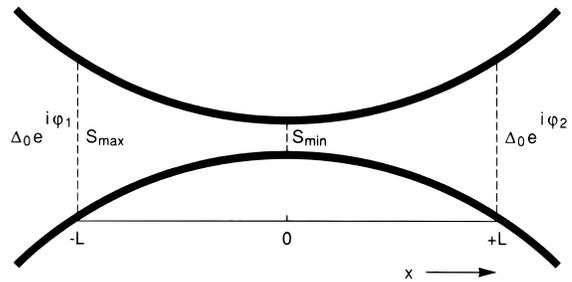}}
\caption{
Schematic drawing
of a superconducting constriction of slowly varying width. (Taken from
Ref.\ \protect\cite{Bee91}.)
\label{fig2}
}
\end{figure}

Let us now consider an impurity-free superconducting constriction
(Fig.\ 2), whose cross-sectional area
$S(x)$ widens from $S_{\rm min}$ to $S_{\rm max}\gg S_{\rm min}$ over a
length $L\gg\lambda_{F}$ and $\mbox{}\ll\xi_{0}$. The $x$-axis is along
the constriction, and the area $S_{\rm max}$ is reached at $x=\pm L$.
We are interested in the case that only a small number $N\simeq S_{\rm
min}/\lambda_{\rm F}^{2}$ of transverse modes can propagate through the
constriction at energy $E_{\rm F}$, and we assume that the propagation
is {\em adiabatic}, i.e.\ without scattering between the modes. In the
superconducting reservoirs to the left and right of the constriction
the pair potential (of absolute value $\Delta_{0}$) has phase
$\phi_{1}$ and $\phi_{2}$, respectively. We wish to calculate the
current $I(\delta\phi)$ which flows in equilibrium through the
constriction, for a given (time-independent) phase difference
$\delta\phi\equiv\phi_{\rm 1}-\phi_{\rm 2}\in(-\pi,\pi)$ between the
two reservoirs.  The characterization of the reservoirs by a constant
phase is not strictly correct. The phase of the pair potential has a
gradient in the bulk if a current flows. The gradient is $1/\xi_{0}$
when the current density equals the critical current density in the
bulk. In our case the critical current is limited by the point contact,
so that the gradient of the phase in the bulk is much smaller than
$1/\xi_{0}$ (by a factor $S_{\rm min}/S(x)\ll 1$). Since the excitation
spectrum is determined by the region within $\xi_{0}$ from the
junction, one can safely neglect this gradient in calculating
$I(\delta\phi)$ from Eq.\ (\ref{dFdphi2}).

As we enter the constriction from, say, the left reservoir, the pair
potential $\Delta({\bf r})$ begins to vary from the bulk value
$\Delta_{0}\exp({\rm i}\phi_{1})$, for two reasons: ({\em a}) because
the small number of transverse modes leads to a non-uniform density;
and ({\em b}) because of contributions from the $N$ modes which have
reached ${\bf r}$ from the right reservoir. Non-uniformities in
$\Delta$ due to ({\em a}) and ({\em b}) are of order $1/N(x)\simeq
\lambda_{\rm F}^{2}/S(x)$ and $N/N(x)\simeq S_{\rm min}/S(x)$,
respectively ($N(x)$ being the number of propagating modes at $x$). For
$|x|>L$ we have both $S(x)\gg\lambda_{\rm F}^{2}$ and $S(x)\gg S_{\rm
min}$. One may therefore neglect these non-uniformities for $|x|>L$,
and put \begin{eqnarray} \Delta({\bf r})=\left\{\begin{array}{ll}
\Delta_{0}{\rm e}^{{\rm i}\phi_{1}}&\;{\rm if}\;x<-L,\\ \Delta_{0}{\rm
e}^{{\rm i}\phi_{2}}&\;{\rm if}\;x>L.  \end{array}\right.\label{Delta0}
\end{eqnarray} Note the difference with planar SNS junctions, where
$\Delta$ recovers its bulk value only at a distance $\xi_{0}$ from the
SN interface. The much shorter decay length for non-uniformities due to
a constriction is a geometrical ``dilution'' effect, well known in the
theory of weak links \cite{Lik79,Kul77}. No specific assumptions are
made on the variation of $\Delta$ in the narrow part of the
constriction. In particular, our analysis also applies to a
non-superconducting constriction ($\Delta({\bf r}) =0$ at $S_{\rm
min}$). Eq.\ (\ref{Delta0}) remains valid up to terms of order $S_{\rm
min}/S_{\rm max}\ll 1$.

We describe the propagation of quasiparticles through the constriction
by means of the BdG equations (\ref{BdG1}). The electrostatic potential
$V({\bf r})$ is now the confining potential which defines the
constriction. We neglect the vector potential induced by the Josephson
current. This is allowed if the total flux penetrating the junction is
small compared to the flux quantum $h/2e$, which is generally the case
if the constriction is small compared to $\xi_{0}$.\footnote{More
precisely, the constriction should be small compared to the penetration
depth of the magnetic field in the Josephson junction, which length is
much larger than $\xi_{0}$ in a type II superconductor \cite{Lik79}.}
For ${\bf A}({\bf r})\equiv 0$, the transverse modes
$|n\rangle\equiv\phi_{n}({\bf r})$ are eigenfunctions of
$(p_{y}^{2}+p_{z}^{2})/2m+V({\bf r})$, with eigenvalues $E_{n}(x)$. We
expand $\Psi({\bf r})=\sum_{n}({\rm u}_{n},{\rm v}_{n})\phi_{n}$ into
transverse modes and neglect off-diagonal matrix elements $\langle
n|{\cal H}|n'\rangle$ and $\langle n|\Delta|n'\rangle$
($\langle\ldots\rangle$ denotes integration over $y$ and $z$). This is
the adiabatic approximation. The functions ${\rm u}_{n} (x)$ and ${\rm
v}_{n}(x)$ then satisfy the one-dimensional BdG equations
\begin{eqnarray} (p_{x}^{2}/2m-U_{n}){\rm u}_{n}+\Delta_{n}{\rm
v}_{n}=\epsilon {\rm u}_{n},\nonumber\\ -(p_{x}^{2}/2m-U_{n}){\rm
v}_{n}+\Delta_{n}^{\ast} {\rm u}_{n}=\epsilon{\rm v}_{n},\label{BdG2}
\end{eqnarray} where $U_{n}(x)=E_{\rm F}-E_{n}(x)-(1/2m)\langle
n|p_{x}^{2}| n\rangle\approx E_{\rm F}-E_{n}(x)$ is the kinetic energy
of motion along the constriction in mode $n$, and
$\Delta_{n}(x)=\langle n|\Delta|n\rangle$ is the projection of
$\Delta({\bf r})$ onto the $n$-th mode. We will consider one mode
$n\leq N$ at a time, and omit the subscript $n$ for notational
simplicity in most of the equations.

In order to simplify the solution of the 1D BdG equations (\ref{BdG2}),
we adopt the WKB method of Bardeen et al. \cite{Bar69}, which consists
in substituting \begin{eqnarray} \left(\begin{array}{c}{\rm u}\\{\rm v}
\end{array}\right)= \left(\begin{array}{c}{\rm e}^{{\rm i}\eta/2}
\\{\rm e}^{-{\rm i}\eta/2} \end{array}\right)\exp\left( {\rm
i}\int_{0}^{x}k(x')dx' \right) \label{WKB1} \end{eqnarray} into
Eq.\ (\ref{BdG2}) and neglecting second order derivatives (or products
of first order derivatives). One thus generalizes the familiar WKB
method for the Schr\"{o}dinger equation to the BdG equations, by having
not only a spatially dependent wavevector $k(x)$---but also spatially
dependent coherence factors $\exp[\pm{\rm i}\eta(x)/2]$. The resulting
equations for $\eta(x)$ and $k(x)$ are \begin{eqnarray}
-(\hbar^{2}/2m)k\eta'+\epsilon&=&|\Delta|\cos(\eta-\phi), \label{eta}\\
(\hbar^{2}/2m)(k^{2}-{\rm i}k')-U&=& {\rm
i}|\Delta|\sin(\eta-\phi),\label{k} \end{eqnarray} where
$\Delta(x)\equiv |\Delta(x)|{\rm e}^{{\rm i}\phi(x)}$. In general, both
$\eta$ and $k$ are complex. The WKB approximation requires that $U$
changes slowly on the scale of $\lambda_{\rm F}$, so that reflections
due to abrupt variations in the confining potential can be neglected.
Reflections (accompanied by a change in sign of ${\rm Re}\;k$) due to
spatial variations in the pair potential are negligible provided that
$|\Delta|$ is much smaller than the kinetic energy $U$ of motion along
the constriction.  Since $U\gtrsim E_{\rm F}-E_{N}(0)$, the WKB method
can not treat the threshold regime that $E_{\rm F}$ lies within
$\Delta_{0}$ from the cutoff energy $E_{N}(0)$ of the highest mode $N$
at the narrowest point of the constriction ($x=0$). The energy
separation $\delta E\equiv E_{N+1}(0)-E_{N}(0)\simeq E_{\rm F}/N$ is
much larger than $\Delta_{0}$ for small $N$, so that the threshold
regime $|E_{\rm F}-E_{N}(0)|\lesssim \Delta_{0}$ consists only of small
intervals in Fermi energy (smaller than the non-threshold intervals by
a factor $\Delta_{0}/\delta E\ll 1$).

For $|x|>L$, where $\Delta$ is independent of $x$, one has a constant
$\eta$ which can take on the two values $\eta^{\rm e}$ and $\eta^{\rm
h}$, \begin{eqnarray} \eta^{\rm e,h}=\phi+\sigma^{\rm e,h}
\arccos(\epsilon/\Delta_{0}),\label{etaeh} \end{eqnarray} where
$\sigma^{\rm e}\equiv 1$, $\sigma^{\rm h}\equiv -1$. We have
$\phi=\phi_{1}$ for $x<-L$ and $\phi=\phi_{2}$ for $x>L$. The function
$\arccos t$ is defined such that $\arccos t\in(0,\pi/2)$ for $0<t<1$;
for $t>1$, one has ${\rm i}\arccos t= \ln[t+(t^{2}-1)^{1/2}]$. The
(unnormalized) WKB wavefunctions $\Psi_{\pm}^{\rm e,h} (x)$ for $|x|>L$
describe an electron-like (e) or hole-like (h) quasiparticle with
positive ($+$) or negative ($-$) wavevector, 
\begin{eqnarray}
\Psi_{\pm}^{\rm e,h}&=&(k^{\rm e,h})^{-1/2}\left( \begin{array}{c} {\rm
e}^{{\rm i}\eta^{\rm e,h}/2} \\{\rm e}^{-{\rm i}\eta^{\rm e,h}/2}
\end{array} \right)\exp\left( \pm{\rm i}\int_{0}^{x} k^{\rm
e,h}dx'\right) ,\nonumber\\
&&\label{WKB2}\\ k^{\rm
e,h}&=&(2m/\hbar^{2})^{1/2}[U+\sigma^{\rm e,h}
(\epsilon^{2}-\Delta_{0}^{2}) ^{1/2}]^{1/2}.\label{keh} 
\end{eqnarray}
The square roots are to be taken such that ${\rm Re}\; k^{\rm e,h}\geq
0$, ${\rm Im}\;k^{\rm e}\geq 0$, ${\rm Im}\;k^{\rm h}\leq 0$. The
wavefunction (\ref{WKB2}) is a solution for $|x|>L$ of the BdG
equations up to second order derivatives. One can verify that for zero
confining potential the solution (\ref{WKB2}), (\ref{keh}) is
equivalent to the energy spectrum and coherence factors given in
Eqs.\ (\ref{uniform1})--(\ref{uniform3}).

For $\epsilon>\Delta_{0}$, the wavevectors $k^{\rm e,h}$ are real at
$|x|>L$, and hence the wavefunctions (\ref{WKB2}) remain properly
bounded as $|x|\rightarrow\infty$. This is the continuous spectrum. The
density of states $\rho^{\rm e,h}(\epsilon)$ of the electron and hole
branches of the excitation spectrum is 
\begin{eqnarray} \rho^{\rm
e,h}&=&\frac{1}{2\pi}\,\frac{d}{d\epsilon}\left( \int_{-L_{\rm S}}
^{L_{\rm S}}\!k^{\rm e,h}dx+\int_{-L}^{L}\!\delta k^{\rm e,h}dx \right)
,\label{rho}\\ \delta k^{\rm e,h}&=&(2m/\hbar^{2})^{1/2}[U+\sigma^{\rm
e,h} (\epsilon^{2}-|\Delta|^{2})^{1/2}]^{1/2}-k^{\rm
e,h}.\nonumber\\
&&\label{deltakeh} 
\end{eqnarray} 
The length $L_{\rm S}$ of the
superconducting regions at opposite sides of the constriction is
assumed to be much larger than $\xi_{0}$. The wavevector change $\delta
k^{\rm e,h}$ is of order $1/\xi_{0}$ for $\epsilon\gtrsim\Delta_{0}$
and decays as $\epsilon^{-3/2}$ for $\epsilon\gg\Delta_{0}$.  It
follows that the integral over the continuous spectrum in the
expression (\ref{dFdphi2}) for the Josephson current is of order
$N(e\Delta_{0}/\hbar)L/\xi_{0}$, which vanishes in the limit
$L/\xi_{0}\rightarrow 0$.

For $0<\epsilon<\Delta_{0}$, the wavevectors $k^{\rm
e,h}$ have a non-zero imaginary part at $|x|>L$, so that
Eq.\ (\ref{WKB2}) is not an admissible wavefunction for both $x>L$ and
$x<-L$. The two bounded WKB wavefunctions for $0<\epsilon<\Delta_{0}$
are given at $|x|>L$ by \begin{eqnarray}
\Psi_{+}&=&\left\{\begin{array}{ll} A_{+}\Psi_{+}^{\rm h}&\;{\rm
if}\;x<-L,\\ B_{+}\Psi_{+}^{\rm e}&\;{\rm if}\;x>L,
\end{array}\right.\label{Psip}\\ \Psi_{-}&=&\left\{\begin{array}{ll}
A_{-}\Psi_{-}^{\rm e}&\;{\rm if}\;x<-L,\\ B_{-}\Psi_{-}^{\rm h}&\;{\rm
if}\;x>L.  \end{array}\right.\label{Psim} \end{eqnarray} The transition
from $k^{\rm h}$ to $k^{\rm e}$ on passing through the constriction
(associated with a change in sign of ${\rm Im}\;k$) is analogous to
{\em Andreev reflection} at an SN interface \cite{And64}. Andreev
reflections are to be distinguished from ordinary reflections involving
a change in sign of ${\rm Re}\;k$. Ordinary reflections due to the pair
potential are neglected in the WKB approximation. By matching the
wavefunctions $\Psi_{\pm}$ to the region $|x|<L$ we obtain a
boundary-value problem with a discrete energy spectrum. Since
$\epsilon<\Delta_{0} \ll U$, we may approximate $k\approx
\pm(2mU/\hbar^{2})^{1/2}$ in Eq.\ (\ref{eta}) (the upper sign refers to
$\Psi_{+}$, the lower sign to $\Psi_{-}$). The boundary-value problem
then becomes 
\begin{eqnarray} 
&&\pm[\hbar^{2}U(x)/2m]^{1/2}
\eta'(x)+|\Delta(x)|\cos[\eta(x)-\phi(x)]=\epsilon,\label{bvp}\nonumber\\
&&\\
&&\eta(-L)=\phi_{1}\mp\arccos(\epsilon/\Delta_{0}),\nonumber\\
&&\eta(+L)=\phi_{2}\pm\arccos(\epsilon/\Delta_{0}).\label{bc}
\end{eqnarray} 
Noting that $\eta$ is real, one deduces from
Eq.\ (\ref{bvp}) the inequality
$|\eta'|<(\epsilon+|\Delta|)(\hbar^{2}U/2m)^{-1/2}$. Since
$|\Delta|\lesssim\Delta_{0}$ and $U\gtrsim E_{\rm F}-E_{N}(0)$, we
have the limiting behavior $|\eta(L)-\eta(-L)|
\lesssim(L/\xi_{0})(1-E_{N}(0)/E_{\rm F})^{-1/2}\rightarrow 0$ in the
limit $L/\xi_{0}\rightarrow 0$. Hence, to order $L/\xi_{0}$ the
bound-state energy $\epsilon$ is determined by \begin{eqnarray}
\arccos(\epsilon/\Delta_{0})=\pm{\textstyle\frac
{1}{2}}\delta\phi,\label{ebound} \end{eqnarray} {\em independent of the
precise behavior of $\Delta({\bf r})$ in the constriction}. Since
$\arccos(\epsilon/\Delta_{0})>0$, there is a {\em single\/} bound state
per mode, with energy {\em independent of the mode index\/} $n\leq N$.
This $N$-fold degenerate bound state at energy
$\Delta_{0}\cos(\delta\phi/2)$ differs qualitatively from the Andreev
levels \cite{And64} in an SNS junction with $L_{\rm N}\gg\xi_{0}$,
which are sensitive to the mode index, to the length $L_{\rm N}$ of the
junction, and also to the pair-potential profile at the SN interfaces.
This sensitivity is at the origin of the geometry-dependent critical
current of the SNS junction which we discussed in the Introduction.

We are now ready to calculate the Josephson current from
Eq.\ (\ref{dFdphi2}). (The alternative calculation based on
Eq.\ (\ref{j}) was given in Ref.\ \cite{Bee91}, and leads to the same
final result.) The integral over the continuous part of the excitation
spectrum does not contribute in the limit $L/\xi_{0}\rightarrow 0$ (see
above).  The spatial integral of $|\Delta|^{2}/g$ contributes only over
the region within the constriction ($|\Delta|$ being independent of
$\delta\phi$ in the reservoirs). Since
$|\Delta|^{2}/g\sim\Delta_0/\lambda_{\rm F}\xi_{0}$, one can estimate
\begin{eqnarray} \frac{2e}{\hbar}\int\!d{\bf r}\,|\Delta|^{2}/g\sim
N(e\Delta_{0}/\hbar)\frac{L}{\xi_{0}},\label{estimate} \end{eqnarray}
which vanishes in the limit $L/\xi_{0}\rightarrow 0$. What remains is
the sum over the discrete spectrum in Eq.\ (\ref{dFdphi2}).
Substituting $\epsilon_{p}=\Delta_{0}\cos(\delta\phi/2)$, $p=1,2,\ldots
N$, we obtain the final result for the Josephson current through a
quantum point contact:  
\begin{equation}
I(\delta\phi)=N\frac{e}{\hbar}\Delta_{0}(T)
\sin(\delta\phi/2)\tanh\left(\frac{\Delta_{0}(T)} {2k_{\rm
B}T}\cos(\delta\phi/2)\right).\label{final} 
\end{equation}

Since $N$ is an integer, Eq.\ (\ref{final}) tells us that $I$ for a
given value of $\delta\phi$ increases stepwise as a function of the
width of the constriction. At $T=0$ we have
$I(\delta\phi)=N(e\Delta_{0}/\hbar)\sin(\delta\phi/2)$, with a critical
current $I_{\rm c}=Ne\Delta_{0}/\hbar$ (reached at $\delta\phi=\pi$).
Near the critical temperature $T_{\rm c}$ we have $I(\delta\phi)=
N(e\Delta_{0}^{2}/4\hbar k_{\rm B}T_{\rm c})\sin(\delta\phi)$, and the
critical current is reached at $\delta\phi=\pi/2$. The ratio $I/G$
(with $G=2Ne^{2}/h$ the normal-state conductance of the quantum point
contact) does not contain $N$ and is formally identical to the result
for a classical point contact \cite{Kul77}.\footnote{The identity of
the classical and quantum results for $I/G$ should not be mistaken for
a universal truth: It only holds within the approximations made in our
analysis, and breaks down in particular in the transition region
between two subsequent plateaux of discretized supercurrent.}

\begin{figure}
\centerline{\includegraphics[width=8cm]{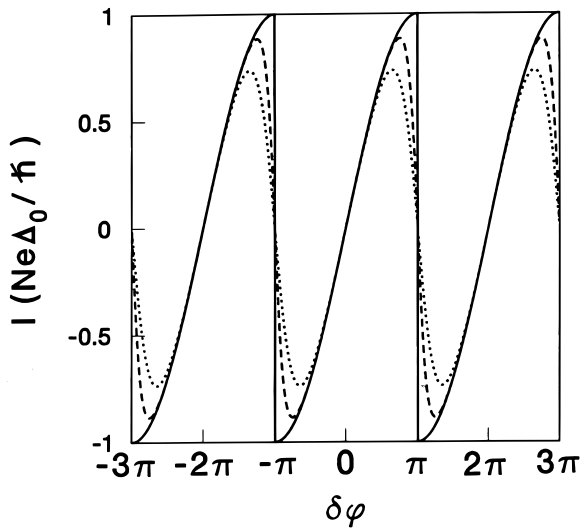}}
\caption{
Current-phase relationship of
a superconducting quantum point contact, according to
Eq.\ (\protect\ref{final}), for three different temperatures (in units
of $\Delta_{0}(0)/k_{\rm B}$): $T=0$ (solid curve), $T=0.1$ (dashed
curve), and $T=0.2$ (dotted curve).
\label{fig3}
}
\end{figure}

In Fig.\ 3 we have
plotted $I(\delta\phi)$ from Eq.\ (\ref{final}) for three different
temperatures: $T=0,0.1,0.2$ in units of $\Delta_{0}(0)/k_{\rm
B}=1.76\,T_{\rm c}$. For these relatively low temperatures the
$T$-dependence of $\Delta_{0}$ may be neglected. The $T$-dependence of
the Josephson current is easily understood within the semiconductor
model (cf.\ Sec.\ II), as a cancellation of the current due to the
states in the energy gap at $-\epsilon_{p}$  and at $+\epsilon_{p}$ .
States above the Fermi energy are unoccupied at $T=0$, but become
occupied when $4k_{\rm B}T\gtrsim\epsilon_{p}$. The supercurrent
decays most rapidly near $|\delta\phi|\simeq\pi$, since $\epsilon_{p}$
is smallest there.

\section{\label{sec4} Experimental realization}

A superconducting quantum point contact (SQPC) of fixed width might be
constructed relatively easily as a constriction or microbridge in a
superconductor of uniform composition. A test of the present theory,
however, requires the fabrication of nanostructures with variable width
or density. In the case of normal-state quantum point contacts this has
been realized by depleting the high-mobility two-dimensional electron
gas (2DEG) at the interface of a GaAs-AlGaAs heterostructure by
applying a negative voltage to a split gate on top of the
heterostructure \cite{Wee88,Wha88,Eer91}.  This is an ideal system for
the study of ballistic transport of normal electrons, because of the
long mean free path ($l\simeq 10\,\mu{\rm m}$) and Fermi wave length
($\lambda_{\rm F}\simeq 50\,{\rm nm}$).

An SQPC might be realized in the same system, if superconducting
contacts to the 2DEG can be made \cite{Iva87}. An observation of the
discretized critical current requires that the superconducting regions
extend well into the constriction on either side, as illustrated
schematically in Fig.\ 4 (black regions).  

\begin{figure}
\centerline{\includegraphics[width=8cm]{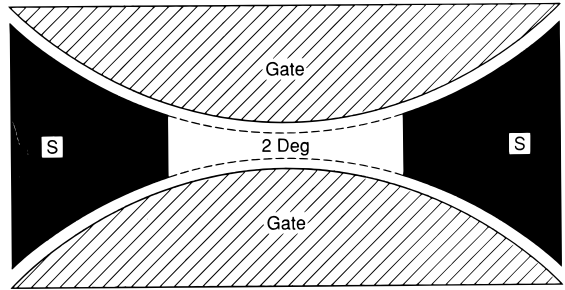}}
\caption{
Top view of a possible experimental realization
of a superconducting quantum point contact.
\label{fig4}
}
\end{figure}

A remaining
problem is the mismatch in Fermi energy that will in general exist at
the interface between the 2DEG (where $E_{\rm F}\simeq 10\,{\rm meV}$)
and the superconductors (where $E_{\rm F}\simeq 1\,{\rm eV}$). This
mismatch, if it is abrupt, induces normal reflections (rather than
Andreev reflections) of quasiparticles incident on the reservoirs.  The
fabrication of sufficiently clean superconducting contacts to the 2DEG
in a GaAs-AlGaAs heterostructure may be difficult, because contacting
requires an alloying process. This problem may be circumvented by using
the surface 2DEG present as a natural inversion layer on p-InAs
\cite{Tak85}. Superconducting contacts, with a shape as in Fig. 4, may
then be evaporated directly onto the surface. Once the contacts have
been made, a constriction of variable width may again be realized by
means of split gates (insulated from the surface 2DEG). Because of the
non-linear dependence of the supercurrent on the gate voltage, such an
SQPC has possible device applications \cite{Hou91}.

\acknowledgments

This research
was supported in part by the National Science Foundation under Grant
No. PHY89--04035.

\end{document}